\def\CMP{\sevenrm Commun.\ Math.\ Phys.}
\def\JMP{\sevenrm J.\ Math.\ Phys.}

\def\LMP{\sevenrm Lett.\ Math.\ Phys.}

%
%            DATUM
%
\def\today{\number\day .\space\ifcase\month\or
January\or February\or March\or April\or May\or June\or
July\or August\or September\or October\or November\or December\fi, \number \year}
%
%           Theorem, Proposition, Corollary, Lemma, Conjecture
%
\newcount \theoremnumber
\def\cleartheoremnumber{\theoremnumber = 0 \relax}

\def\Prop #1 {
             \advance \theoremnumber by 1
             \vskip .6cm 
             \goodbreak 
             \noindent
             {\bf Proposition {\the\headlinenumber}.{\the\theoremnumber}.}
             {\sl #1}  \goodbreak \vskip.8cm}

\def\Conj#1 {
             \advance \theoremnumber by 1
             \vskip .6cm  
             \goodbreak 
             \noindent
             {\bf Conjecture {\the\headlinenumber}.{\the\theoremnumber}.}
             {\sl #1}  \goodbreak \vskip.8cm} 

\def\Th#1 {
             \advance \theoremnumber by 1
             \vskip .6cm  
             \goodbreak 
             \noindent
             {\bf Theorem {\the\headlinenumber}.{\the\theoremnumber}.}
             {\sl #1}  \goodbreak \vskip.8cm}

\def\Lm#1 {
             \advance \theoremnumber by 1
             \vskip .6cm  
             \goodbreak 
             \noindent
             {\bf Lemma {\the\headlinenumber}.{\the\theoremnumber}.}
             {\sl #1}  \goodbreak \vskip.8cm}

\def\Cor#1 {
             \advance \theoremnumber by 1
             \vskip .6cm  
             \goodbreak 
             \noindent
             {\bf Corollary {\the\headlinenumber}.{\the\theoremnumber}.}
             {\sl #1}  \goodbreak \vskip.8cm} 
%
%            Formelnummerierung
%
\newcount \equationnumber

\newcount \refnumber

\def\[]    {\global 
            \advance \refnumber by 1
            [{\the\refnumber}]}

\def\# #1  {\global 
            \advance \equationnumber by 1
            $$ #1 \eqno ({\the\equationnumber}) $$ }

\def\% #1 { \global
            \advance \equationnumber by 1
            $$ \displaylines{ #1 \hfill \llap ({\the\equationnumber}) \cr}$$} 

\def\& #1 { \global
            \advance \equationnumber by 1
            $$ \eqalignno{ #1 & ({\the\equationnumber}) \cr}$$}
%
%                             Referencen
%                                
\newcount \Refnumber

\def\Ref #1 #2 #3 #4 #5 #6  {\ninerm \global
                             \advance \Refnumber by 1
                             {\ninerm #1,} 
                             {\ninesl #2,} 
                             {\ninerm #3.} 
                             {\ninebf #4,} 
                             {\ninerm #5,} 
                             {\ninerm (#6)}\nobreak} 
\def\Bookk #1 #2 #3 #4       {\ninerm \global
                             \advance \Refnumber by 1
                             {\ninerm #1,}
                             {\ninesl #2,} 
                             {\ninerm #3,} 
                             {(#4)}}
\def\Book{\cr
{\the\Refnumber} &
\Bookk}
\def\Reff{\cr
{\the\Refnumber} &
\Ref}
\def\REF #1 #2 #3 #4 #5 #6 #7   {{\sevenbf [#1]}  & \hskip -9.5cm \vtop {
                                {\sevenrm #2,} 
                                {\sevensl #3,} 
                                {\sevenrm #4} 
                                {\sevenbf #5,} 
                                {\sevenrm #6} 
                                {\sevenrm (#7)}}\cr}
\def\BOOK #1 #2 #3 #4  #5   {{\sevenbf [#1]}  & \hskip -9.5cm \vtop {
                             {\sevenrm #2,}
                             {\sevensl #3,} 
                             {\sevenrm #4,} 
                             {\sevenrm #5.}}\cr}
\def\HEP #1 #2 #3 #4     {{\sevenbf [#1]}  & \hskip -9.5cm \vtop {
                             {\sevenrm #2,}
                             {\sevensl #3,} 
                             {\sevenrm #4.}}\cr}
%
%                               Beweis
%
\def\bull{$\sqcup \kern -0.645em \sqcap$}
%
%               Definition, Remark, Example, Proof  
%
\def\Def#1{  \advance \theoremnumber by 1
             \vskip .6cm  
             \goodbreak 
             \noindent
             {\bf Definition {\the\headlinenumber}.{\the\theoremnumber}.}
             {\sl #1}  \goodbreak \vskip.4cm}
\def\Rem#1{\vskip .4cm \goodbreak \noindent
                                     {\it Remark.} #1 \goodbreak \vskip.5cm }

\def\Pr#1{\goodbreak \noindent {\it Proof.} #1 \hfill \bull  \goodbreak \vskip.5cm}

%
%                               Abstand
%
\def\*{\vskip 1.0cm}      

%
%            †berschriften
%
\newcount \ssubheadlinenumber

\def\SSHL #1 {\goodbreak
            \cleartheoremnumber
            \vskip 1cm
            \advance \ssubheadlinenumber by 1
{\rm \noindent {\the\headlinenumber}.{\the\subheadlinenumber}.{\the\ssubheadlinenumber}. #1}
            \nobreak \vskip.8cm \rm \noindent}
\newcount \subheadlinenumber
\def\clearsubheadlinenumber{\subheadlinenumber = 0 \relax}
\def\SHL #1 {\goodbreak
            \cleartheoremnumber
            \vskip 1cm
            \advance \subheadlinenumber by 1
            {\rm \noindent {\the\headlinenumber}.{\the\subheadlinenumber}. #1}
            \nobreak \vskip.8cm \rm \noindent}
\newcount \headlinenumber

\newcount \headlinesubnumber
\def\clearheadlinesubnumber{\headlinesubnumber = 0 \relax}
\def\Hl #1 {\goodbreak
            \cleartheoremnumber
            \clearheadlinesubnumber
            \clearsubheadlinenumber
            \advance \headlinenumber by 1
            {\bf \noindent {\the\headlinenumber}. #1}
            \nobreak \vskip.4cm \rm \noindent}

\font\twentyrm=cmr17
\font\fourteenrm=cmr10 at 14pt
\font\sevensl=cmsl10 at 7pt
\font\sevenit=cmti7 

\font\css=cmss10
\font\Rosch=cmr10 at 9.85pt
\font\Cosch=cmss12 at 9.5pt
\font\rosch=cmr10 at 7.00pt
\font\cosch=cmss12 at 7.00pt
\font\nosch=cmr10 at 7.00pt
%
%
%
%
%
%
%
%                   Symbols
%
%
%
%                              
%
%
%
%               reelle, nat\"urliche, ganze, komplexe Zahlen
%
%    Beispiel:  ${\bf E}_\alpha {s \atop {\raise 4pt \hbox{$\to$}}} 1$
%
\def\Z                 {\hbox{{\css Z}  \kern -1.1em {\css Z} \kern -.2em }}
\def\R                 {\hbox{\raise .03ex \hbox{\Rosch I} \kern -.55em {\rm R}}}
\def\N                 {\hbox{\rm I \kern -.55em N}}
\def\C                 {\hbox{\kern .20em \raise .03ex \hbox{\Cosch I} \kern -.80em {\rm C}}}

\def\r                 {\hbox{\raise .03ex \hbox{\rosch I} \kern -.45em \hbox{\rosch R}}}
\def\n                 {\hbox{\hbox{\rosch I} \kern -.45em \hbox{\nosch N}}}
\def\c                 {\hbox{\raise .03ex \hbox{\cosch I} \kern -.70em \hbox{\rosch C}}}

\def\z                 {\hbox{\kern 0.2em {\cal z}  \kern -0.6em {\cal z} \kern -0.3em  }}
\def\1                 {\hbox{\rm \thinspace \thinspace \thinspace \thinspace
                                  \kern -.50em  l \kern -.85em 1}}
\def\unit                 {\hbox{\sevenrm \thinspace \thinspace \thinspace \thinspace
                                  \kern -.50em  l \kern -.85em 1}}
%
%
%
%
%
%                   
%                              Sonderzeichen alg. QFT
%
%
%
%                              
%
%

\def\A                 {{\cal A}}

\def\B                 {{\cal B}} 

\def\H                 {{\cal H}} 
 
\def\O                 {{\cal O}}

%

%
%                                
%
%
%                             Referencen
%                                
%
%

%\Z  
%\R  
%\N  
%\C  
%\Ro 
%\No 
%\Zo 
%\Co 
%\r  
%\n  
%\c  
%\ro 
%\no 
%\co 
%\z  
%\1  
%\unit  

\def\versuch #1 #2 {
\vskip -.1 cm
\global \advance \equationnumber by 1
            $$\displaylines{ \rlap{ #1 } \hfill #2  \hfill \llap{({\the\equationnumber})} } $$ 
\vskip  .1cm
\noindent}

\nopagenumbers
\def\Draft  {\hbox{Preprint \today}}
\def\firstheadline{\hss \hfill  \Draft  \hss} 
\headline={
\ifnum\pageno=1 \firstheadline
\else 
\ifodd\pageno \rightheadline 
\else \leftheadline \fi \fi}
\def\rightheadline{\sevenrm THE REEH--SCHLIEDER PROPERTY FOR GROUND STATES
\hfill \folio } 
\def\leftheadline{\sevenrm \folio \hfill CHRISTIAN D.\ J\"AKEL}
\voffset=2\baselineskip
\magnification=1200
%**************************************************************************************************
%
%
%                  TITELSEITE
%
%
%
%**************************************************************************************************

\vskip 1cm

\noindent
{\twentyrm The Reeh--Schlieder Property for Ground States}

\vskip 1cm
\noindent
{\sevenrm CHRISTIAN D.\ J\"AKEL}

\noindent
{\sevenit  Inst.\ f\"ur theoretische Physik,
Technikerstr.\ 25, Universit\"at Innsbruck, Austria}

\noindent
{\sevenit e-mail: christian.jaekel@uibk.ac.at}

\vskip .5cm     
\noindent {\sevenbf Abstract}. {\sevenrm Recently it has been shown that 
the Reeh--Schlieder property w.r.t.\ thermal equilibrium states is a direct consequence of 
locality, additivity and the relativistic KMS condition. Here we extend this result to
ground states.}

\vskip 1 cm

%**************************************************************************************************
%
%
%                  Einleitung
%
%
%
%**************************************************************************************************

\Hl{Introduction}

\noindent
The general theory of quantized fields is mostly concerned with the vacuum
sector, since this is the appropriate framework for traditional (i.e., few particle)
high energy physics. However, 
speculations about a phase-transition (at high temperatures and high densities) 
from standard hadronic matter to a state which is commonly called the 
quark--gluon plasma activated some interest in thermal sectors. Recent experiments on 
cold Bose and Fermi gases have renewed the interest in ground states or
(low temperature) thermal states. Just as in the vacuum sector we suppose 
that some of the most peculiar predictions of the theory will be traced back to the 
famous Reeh--Schlieder property [RS], altough this may take some time and effort.
 For instance, the significance of the Reeh--Schlieder 
property for quantum information theory was realized only recently (see, e.g., [CH][SW]).) 
Despite its direct physical significance, the Reeh--Schlieder 
property will turn out be indispensable as a technical tool, e.g., for
scattering or superselection theory in the new sectors.

According to the standard arguments the Reeh--Schlieder property
is a property of finite energy states in the vacuum sector.
One might argue that all physically 
relevant states should be locally normal w.r.t.\ the vacuum representation and therefore, 
whenever the Reeh--Schlieder property is urgently needed, one may take recourse to 
the vacuum sector. But obviously it is more enlightening to prove the Reeh--Schlieder 
property directly in the sector that is under investigation (see, e.g., [J\"a][Ju] for
KMS states). Moreover, it is not always possible to take recourse to the vacuum sector.
In lower space--time dimensions KMS states, which fail to be locally normal, 
exist (see, e.g., the discussion in [BJu]). 
Despite the general belief that this can not happen in 3+1 
space--time dimensions, all attempts to rule out infrared problems, which may destroy 
local normality, failed up to now. 

From a technical viewpoint we would like to emphasize that the original arguments
of Reeh and Schlie\-der [RS] were based on the global symmetry properties of the vacuum state. 
Thus the real challenge may come from quantum field theory on curved space--times (see
[St] and [V] for free fields) where the curved background will, at least in general, not 
allow global symmetries. (See however, [BoB] and [BEM] for highly symmetric space--times.)
Thus the question arises whether one can abandon the assumption that the translations are 
unitarily implemented. And indeed,
it has been shown that the Reeh--Schlieder property w.r.t.\ thermal equilibrium states is a 
direct consequence of locality, additivity and the relativistic KMS condition of 
Bros and Buchholz [BB] which does not exclude KMS states braking the 
rotation or translation symmetry [J\"a].
In this letter we extend this result to a class of ground states 
(in Minkowski space, of course), which satisfy a similar `relativistic ground state 
condition'. Once again the essential steps of our proof are based on a theorem of 
Glaser. They exploit only the characteristic analyticity properties
of a relativistic ground state;
whether or not the translation or/and rotation symmetries 
are sponanteously broken turns out to be 
irrelevant for the Reeh--Schlieder property. From a technical 
viewpoint the original contribution of this letter is contained in Lemma 3.4. 

To conclude this introduction, we briefly outline the content of this letter. In Section~2
we introduce a (relativistic) ground state condition and discuss some aspects of the 
corresponding GNS representations.
Section 3 contains the derivation of the Reeh--Schlieder property for ground 
states.

\vskip 1cm

\Hl{Relativistic Ground States}

\noindent
In the algebraic formulation [H] a QFT is casted into an inclusion preserving map 
\# {\O \to \A (\O)} 
which assigns to any open bounded region $\O$ in Minkowski 
space $\R^4$ a unital $C^*$-al\-gebra~$\A (\O)$. 
The Hermitian elements of the {\sl abstract} $C^*$-algebra $\A (\O)$ are interpreted as 
the observables which can be measured at times and locations in $\O$. 
The net $\O \to \A(\O)$ is isotonous, i.e., there exists a unital embedding
\# {\A(\O_1) \hookrightarrow \A(\O_2)
\qquad \hbox{if} \quad \O_1 \subset \O_2.} 
For mathematical convenience the local algebras are embedded in the  $C^*$-inductive limit
algebra 
\# { \A = \overline{ \cup_{ {\cal O}  \subset \r^4} \A(\O) }^{\, C^*}  .}  
The space--time symmetry of Minkowski space manifests itself
in the existence of a representation 
\# {\alpha \colon ( \Lambda, x) \mapsto \alpha_{ \Lambda, x} \in Aut (\A), 
\qquad  (\Lambda, x) \in {\cal P}_+^\uparrow ,} 
of the (orthochronous) Poincar\'e group ${\cal P}_+^\uparrow$. 
Lorentz transformations $\Lambda$  and space--time translations~$x$ act geometrically:
\# {\alpha_{ \Lambda, x} \bigl( \A (\O) \bigr) 
= \A (\Lambda \O + x)  \qquad \forall (\Lambda, x) \in {\cal P}_+^\uparrow.} 
Without loss of generality, we may assume that the space--time translations
$\alpha \colon \R^4 \to Aut (\A)$ are strongly continuous, i.e., for each $ a \in \A$
\# { \| \alpha_x (a) - a \| \to 0 \qquad \hbox{as} \qquad x \to 0.}
Observables localized in spacelike separated space--time regions commute, i.e.,
\# { \A (\O_1) \subset \A^c ( \O_2) \quad \hbox{\rm if} \quad \O_1 \subset \O_2'.}
Here $\O '$ denotes the spacelike complement of $\O$ and 
$\A^c (\O)$ denotes the set of operators in~$\A$ which commute with all operators in $\A(\O)$.

\vskip 1cm

States are, by definition, positive, linear and normalized functionals over $\A$. 
If stable crystals exist in a relativistic framework, then they will certainly 
break the spatial translation and rotation symmetry of space and time. Consequently,
one should not (and we do not) require that these symmetries can be unitarily implemented
in the GNS representation associated with such a state.  
The maximal propagation velocity of signals, however, will not be affected by such a 
lack of symmetry. It is simply characteristic for any relativistic theory. Following 
Bros and Buchholz [BB] we propose that it manifests itself in the following 
`relativistic ground state condition'.

\Def{A time invariant state $\omega_\infty $ is a relativistic ground state  if and only
if for every pair of elements $a, b$ of $\A$ there exists a function~${\cal F}_{a,b}$ which is
bounded and analytic in a convex open tube
\# { -{\cal T}  \times {\cal T} , \qquad {\cal T} \subset  \{ z \in \C : \Im z \in V_+  \}.}
where the basis ${\cal C}$ of ${\cal T} = \R^4 +i {\cal C}$ is a neighbourhood in $\R^4$ of the linear segment 
\# { \{ y \in \R^4 : y = \lambda e , \lambda > 0 \},}
$e$ is a timelike unit vector and at all boundary points $x$ in $\R^4$ the cone $\Lambda_x$ with apex in $x$,
which is the union of all closed half-lines starting from $x$ and intersecting ${\cal C}$,
is the light cone $V_+ = \{ y \in \R^4 : y^0 >  | \vec y | \}$.
Moreover, ${\cal F}_{a,b}$ is continuous at the boundary set $\R^4 \times \R^4$ with boundary values given by
\# 
{ {\cal F}_{a,b} (x_1, x_2)   = \omega_\infty \bigl( \alpha_{x_1} ( a) \alpha_{x_2} (b) \bigr)  
\qquad \forall x_1, x_2 \in \R^4 .}  
}

% \Rem{We recall: Let be a point $b$ in $\R^4$ which belongs to the boundary of ${\cal C}$.
% At any point $y= x + ib$ of its boundary the cone $\Lambda_b$ with apex $b$ in $\R^4$ 
% which is the union of all closed half-lines starting from $b$ and intersectin 
% ${\cal C}$ is called the profile of the tube 
% ${\cal C}$.} 

Obviously (see condition (ii) in the following theorem), relativistic ground states are   
ground states in the usual sense. Recall [BR, 5.3.19] the following

\Th{A state $\omega_\infty$ is called a {\it ground state} if it satisfies one (and thus all)
of the following four equivalent conditions w.r.t.\ some unit vector $e$ in the 
forward light-cone $V_+$:
\vskip .3cm
\halign{ \indent #  \hfil & \vtop { \parindent =0pt \hsize=34.8em
                            \strut # \strut} \cr 
(i)     & If $a,b \in \A_{\alpha_e}$, then the entire analytic function 
\# { z \mapsto \omega_\infty \bigl( a \alpha_{\tau e} (b) \bigr) }
is uniformly bounded in the region $\{ \tau \in \C : \Im \tau \ge 0 \}$.
${\A}_{\alpha_e} \subset \A$
denotes the set of analytic elements for the one-parameter subgroup $t \mapsto \alpha_{te}$.
\cr
(ii)     & For any $a , b \in \A$ there exists a function $F_{a,b}$ which is continuous
in $\Im z \ge 0$ and analytic and bounded in $\Im z >0$. Moreover,  
\# { F_{a,b} (t) = \omega_\infty \bigl( \alpha_{-te/2} (a) \alpha_{te/2} (b) \bigr) 
\qquad \forall t \in \R. }
\cr
(iii)     & Let ${\cal D}$ denote the set of infinitely differentiable 
functions with compact support. If $f$ is a function with Fourier transform
$\tilde f \in {\cal D}$ and supp $\tilde f \subset (- \infty, 0)$,
then
\# { \omega_\infty \bigl( \alpha_{f} (a)^* \alpha_{f}  (a) \bigr) = 0 \qquad \forall a \in \A. }
Here $\alpha_{f} (a):= \int {\rm d}t \, \, f(t) \alpha_{te} (a)$.
\cr
(iv)     & $\omega_\infty$ is time invariant, and if, in the
GNS representation $(\pi_\infty, \H_\infty, \Omega_\infty)$, 
\# { {\rm e}^{it H_\infty } \pi_\infty (a) \Omega_\infty 
= \pi_\infty \bigl( \alpha_{te} (a) \bigr)) \Omega_\infty }
is the corresponding unitary representation of the time evolution $t \mapsto \alpha_{te} $ 
on $\H_\infty$, then
\# { H_\infty \ge 0 .}
\cr}
\noindent
\vskip -.2cm
If these conditions are satisfied, then $U(t) := {\rm e}^{it H_\infty } \in \pi_\infty (\A)''$
for all $ t \in \R$. Note that there might be a distinguished time direction $e$; 
if there is none (as it is the case for the vacuum),
then any timelike unit vector can be used.}
 
Obviously, the set of ground states $K_\infty$ is a weak$^*$-closed convex 
subset of the state space. We recall that the decompostion of a ground state 
into extremal ground states 
is in general not unique. (Note that the KMS states for a fixed temperature 
form a simplex, thus the decompostion of KMS states into extremal ones is always unique.) 
But the decompostion of ground states shows another simple geometric property not 
generally shared by the set of KMS states.
The ground states form a face, i.e., if a ground state
\# { \omega_\infty = \sum_{i =1}^n \lambda_i \omega_i } 
is a finite convex combination of arbitrary states then each $\omega_i$, $i = 1, \ldots , n $,
is automatically a ground state, too.

If $\omega \in K_\infty$ is an extremal ground state, 
then $\omega$ is pure, i.e.,
\# { \pi_\omega (\A)'' = \B(\H_\omega) \qquad \hbox{and} \qquad \pi_\omega (\A)' 
= \C \cdot \1 .}
If the pair $(\A, \omega)$ is $\R$-abelian, i.e.,
\# { \inf_{ a' \in C_\circ (\alpha_{\r e} (a))}
\bigl| \omega' ([ a' , b] )  \bigr| = 0}
for all $a, b \in \A$ and all  time invariant vector states $\omega'$ of $\pi_\omega$,
then $\pi_\omega (\A)'$ is (at least) abelian. Here $C_\circ \bigl( \alpha_{\r e} (a) \bigr) $ 
denotes the convex hull of $\{ \alpha_{te} (a) : t \in \R \}$. 

Let us recall some more well known properties (taken from [BR, 5.3.40]):

\Th{Let $K_\infty$ be the set of ground states.
The following statements are equivalent:
\vskip .3cm
\halign{ \indent #  \hfil & \vtop { \parindent =0pt \hsize=34.8em
                            \strut # \strut} \cr 
(i)     & The pair $(\A, \omega)$ is $\R$-abelian for all $\omega \in K_\infty$.
\cr
(ii)     & $\pi_\omega (\A)'$ is abelian for all $\omega \in K_\infty$.
\cr
(iii)     & $K_\infty$ is a simplex; i.e., there exists a unique decompostion into extremal 
ground states.
\cr
(iv)     & Each pure ground state is weakly clustering in the sense that
\# { \inf_{ a' \in C_\circ (\alpha_{\r e} (a))}
\bigl| \omega (a' b) -\omega (a) \omega (b) \bigr| = 0}
for all $a, b \in \A$.
\cr
(v)     & If $\omega \in K_\infty$ is a factor state, then 
$\omega$ is pure.
\cr
(vi)     & If $\omega_1, \omega_2 \in K_\infty$ are pure states, then
$\omega_1$ and $\omega_2$ are either disjoint
or equal.
\cr
(vii)     & If $\omega_1$ and $\omega_2$ are distinct pure states in $K_\infty$, then the face
generated by 
$\omega_1$ and $\omega_2$ in the set of time invariant states is equal to the convex set
\# { \Bigl\{ \lambda \omega_1 + (1 - \lambda) \omega_2 : \lambda \in [0,1] \Bigr\}.}
\cr
}
}

Let us now return to relativistic ground states.
The GNS representation $\pi_\infty$ assigns to any $\O \subset \R^4$ a von Neumann algebra
\# 
{ \qquad {\cal R}_\infty (\O) = \pi_\infty \bigl( \A(\O) \bigr)''. }
%
% From the experience with certain spin models, one would
% expect that ${\cal R}_\infty := \pi_\infty \bigl( \A(\O) \bigr)''$ is generically of type I;
% we will however not need this assumption in the sequel. 

\Def{The net $\O \to {\cal R}_\infty (\O)$ is called {\it additive},  
if
\# { \cup_{i \in I} \O_i = \O \Rightarrow \vee_{i \in I} {\cal R}_\infty (\O_i) 
= {\cal R}_\infty (\O)  .}
Here $I$ is some index set and $\vee_i {\cal R}_\infty (\O_i)$ denotes the von Neumann  
algebra generated by the algebras ${\cal R}_\infty (\O_i)$, $i \in I$.} 

\Rem{If $\omega_\infty$ is locally normal w.r.t.\ the vacuum representation, then
additivity in the vacuum sector and additivity in the ground state sector are
equivalent. As is well known, additivity in the vacuum sector can be proven, 
if the net of local algebras is constructed from a Wightman field theory.}

\vskip 1cm

\Hl{The Reeh--Schlieder Property}

\noindent
We start with the following

\Prop{Let $\omega_\infty$ be a state which satisfies the {\it relativistic ground state condition}   
and let ${\cal V}$ be an open neighborhood  of the origin in $\R^4$. 
It follows that for each complex-valued test function $f$ with support in ${\cal V}$
\# { \int_{\r^4 \times \r^4} {\rm d}^4 y_1 {\rm d}^4 y_2 \, \, 
{\cal F}_{a^*,a} (y_1 - i \kappa e, y_2 + i  \kappa e) \overline{ f (y_1)} \, f(y_2)  \ge 0 }    
for all $\kappa  > 0$. Here $e$ denotes the unit vector introduced in Theorem 2.2, 
which might not be unique.}

\Pr{Let $a \in \A_\alpha$ be an entire analytic element for the translations. Put 
\# { \Psi_f := \int_{\cal V} {\rm d}^4 y_1 \, \, 
f(y_1) \alpha_{y_1} \bigl(\alpha_{i  \kappa e} (a) \bigr) 
\Omega_\infty \in \H_\infty. }
Exploring the definiton (10) of ${\cal F}_{a^*,a}$ one finds
\# { \int_{\r^4 \times \r^4} {\rm d}^4 y_1 {\rm d}^4 y_2 \, \, 
{\cal F}_{a^*,a} (y_1 - i \kappa e,  y_2 + i  \kappa e) \overline { f (y_1) } f(y_2)  
= \| \Psi_f \|^2 \ge 0.}    
For general $a \in \A$, choose a sequence $\{ a_n \in \A_\alpha \}_{n \in \n}$  such that
\# { \| a_n \| \le \| a \| \qquad \hbox {and} \qquad \pi_\infty (a_n) \Omega_\infty \to 
\pi_\infty (a) \Omega_\infty \quad \hbox {as} \quad n \to \infty.}
Now define, for $y_1, y_2 \in \R^4$ and $\kappa > 0 $,
\# {{\cal F}_n (y_1 - i \kappa e,  y_2 + i  \kappa e) 
:= {\cal F}_{a_n^*,a_n} (y_1 - i \kappa e,  y_2 + i  \kappa e) .}
The maximum modulus principle [R] implies that 
\# { \Bigl| {\cal F}_n (y_1 - i \kappa e,  y_2 + i  \kappa e) -  
{\cal F}_m (y_1 - i \kappa e,  y_2 + i  \kappa e) \Bigr| }
assumes its maximum value on the boundary of its domain and for $\kappa = 0$, the
boundary value, the relativistic ground state condition yields
\& { \lim_{ \kappa \searrow 0} \Bigl| {\cal F}_n (y_1 & - i \kappa e,  y_2 + i  \kappa e) -  
{\cal F}_m (y_1 - i \kappa e ,  y_2 + i  \kappa e ) \Bigr| 
\cr
& \le \sup_{y_1, y_2 \in \r^4} 
\bigl| \omega_\infty \bigl(\alpha_{y_1} (a^*_n) \alpha_{y_2} (a_n) \bigr) -
\omega_\infty \bigl(\alpha_{y_1} (a^*_n) \alpha_{y_2} (a_m) \bigr) \bigr|
\cr
&
\quad +  \sup_{y_1, y_2 \in \r^4} 
\bigl| \omega_\infty \bigl(\alpha_{y_1} (a^*_n) \alpha_{y_2} (a_m) \bigr) -
\omega_\infty \bigl(\alpha_{y_1} (a^*_m) \alpha_{y_2} (a_m) \bigr) \bigr| 
\cr
& \le   \, \| a \| \sup_{y_2 \in \r^4} 
\bigl\| \pi_\infty \bigl(\alpha_{y_2} (a_n -a_m) \bigr) 
\bigr\|  
+  \, \| a \| \sup_{y_1 \in \r^4} 
\bigl\| \pi_\infty \bigl(\alpha_{y_1} (a_m^* -a_n^*)\bigr) 
\bigr\|  .}
In the last inequality we have used $ \| a_n \| = \| a_n^* \| \le \| a \| $ 
and $\| \Omega_\infty \|=1$.
Continuity of the translations (recall that all automorphisms of a $C^*$-algebra 
are continuous), i.e., 
\# { \lim_{a_m \to a_n} \bigl\| \pi_\infty \bigl(\alpha_{y} (a_m -a_n)\bigr) 
\bigr\| = 0 , }
now implies that 
$\{ {\cal F}_n \}_{n \in \n}$ is a Cauchy sequence uniformly on $\overline {\cal U}$, where
\# { {\cal U} :=  \{ (y_1 - i \kappa e ,  y_2 + i  \kappa e )  : y_1, y_2 \in \R^4,
\kappa > 0 \}   .}
The limit function ${\cal F}_\infty$ is therefore continuous and bounded 
on $\overline {\cal U}$ and analytic
in ${\cal U}$. By construction,
\# { {\cal F}_\infty (y_1, y_2) = {\cal F}_{a^*,a} (y_1,  y_2 ) \qquad \forall y_1, y_2 \in \R^4.}
Thus, due to their analyticity properties, the functions ${\cal F}_\infty$ and 
${\cal F}_{a^*,a}$ must
coincide on ${\cal U} $. It follows that
\& { \int_{\r^4 \times \r^4} {\rm d}^4 y_1 & {\rm d}^4 y_2 \, \, 
{\cal F}_{a^*,a} (y_1 - i \kappa e , y_2 + i  \kappa e ) \overline{ f (y_1) } f(y_2)  =
\cr
& 
\lim_{n \to \infty}  \int_{\r^4 \times \r^4} {\rm d}^4 y_1 {\rm d}^4 y_2 \, \, 
{\cal F}_n (y_1 - i \kappa e ,  y_2 + i  \kappa e ) \overline { f (y_1) }f(y_2)
\ge 0.}    
}

The next step uses an adapted and simplified version of Glaser's Theorem 1 
([G a], see also [G b][BEM]):

\Th{(Glaser): Let $a \in \A$ and let ${\cal F}_{a^*,a}$ denote the function introduced in (10). 
The following properties are equivalent:
\vskip .3cm
\halign{ \indent #  \hfil & \vtop { \parindent =0pt \hsize=34.8em
                            \strut # \strut} \cr 
i.)     & There exists an open neighborhood ${\cal V}$ of $0$ in $\R^4$ 
and a point $z_1 \in {\cal T}$ such that 
$z_1 + {\cal V} \subset  {\cal T}$ and such that 
for each complex-valued testfunction $f$ with support in ${\cal V}$
\# { \int_{\r^4 \times \r^4} {\rm d}^4 y_1 {\rm d}^4 y_2 \, \, 
{\cal F}_{a^*,a} (y_1 + \bar z_1, y_2 + z_1) \overline{ f (y_1) } f(y_2)  \ge 0.}    
\cr
ii.)     & There exists a sequence 
$\bigl\{ f_a^{(n)} \colon {\cal T}  \to \C  \bigr\}_{n \in \n} $ of 
functions holomorphic in ${\cal T}$ such that for $(z_1, z_2) \in -{\cal T} \times {\cal T}$
\# { {\cal F}_{a^*,a} (z_1, z_2) 
= \sum_{n =1}^\infty \overline {  f_a^{(n)} (\bar  z_1)} f_a^{(n)} (z_2)}    
holds in the sense of uniform convergence on every compact subset of 
$-{\cal T}  \times {\cal T}$. \cr}
}

The crucial step in the proof of the Reeh--Schlieder property 
is now summarized in the following

\Th{For each $a \in \A$ the vector valued function $\Phi_a \colon \R^4 \to \H_\infty$,
\# { x \mapsto \pi_\infty \bigl( \alpha_{x} (a) \bigr) \Omega_\infty  }  
can be analytically continued from the real axis into the tube
${\cal T}$
such that it is weakly continuous for $\Im z \searrow 0$. }
 
Since $\Omega_\infty$ need not be separating for
$\pi_\infty (\A)''$, the map $b \mapsto \pi_\infty (b) \Omega_\infty $ 
will in general {\it not} be injective. Hence
we can not immediately apply the arguments used
in the case of KMS states.  However, the following lemma assures that at least 
the map $\pi_\infty (b) \Omega_\infty \mapsto {\cal F}_{b^*,b} (x_1, x_2)$ is 
well-defined.

\Lm{Assume $\pi_\infty (b) \Omega_\infty = 0$. It follows that
\# { {\cal F}_{b^*,b} (z_1, z_2) = 0 \qquad \forall (z_1,z_2) \in -{\cal T} \times {\cal T}.}
}

\vskip -.5cm

\Pr{Since a ground state is required to be time
invariant, the time evolution can always be unitarily implemented in the GNS
representation (see (14)):
\# { U(t) \pi_\infty (b) \Omega_\infty := \pi_\infty \bigl( \alpha_{t e} (b) \bigr)
\Omega_\infty.}
Thus $U(t) \Omega_\infty = \Omega_\infty $ for all $ t \in \R$ and therefore 
$\pi_\infty (b) \Omega_\infty = 0$ implies
\# { \pi_\infty \bigl( \alpha_{t e} (b) \bigr) \Omega_\infty 
= U(t) \pi_\infty (b) U(-t) \Omega_\infty = 0
\qquad  \forall t \in \R.}
Consequently, see (12),
\# { F_{b^*,b} (t + i \kappa) = 0 \qquad \forall  t \in \R,
\quad \forall \kappa > 0 .}
Now let $e$ denote the distinguished unit vector in the time direction. 
(If there is a distinguished time direction; otherwise $e$ can be  
any timelike unit vector.) It follows that
\& { {\cal F}_{b^*,b} \Bigl( - {1 \over 2} ( t  + i \kappa) e,  
{1 \over 2} (t + i \kappa)e \Bigr) 
& = F_{b^*,b} (t + i \kappa) 
\cr
& = 0 \qquad \forall  t \in \R,
\quad \forall \kappa > 0 .}
Moreover, for $z_1 \in {\cal T}$, 
\& {  {\cal F}_{b^*,b} ( \bar  z_1 ,  z_1 )  
& =   
\sum_{n=1}^\infty  \overline {  f_b^{(n)} (z_1)  }
f_b^{(n)}  ( z_1) 
\cr
& =   
\sum_{n=1}^\infty \bigl| f_b^{(n)}   (z_1)   \bigr|^2  } 
is a positve bounded function. Therefore it takes its
minimum on the boundary of ${\cal T}$; but according to  
equation (41) $F_{b^*,b}( \bar  z_1 ,  z_1 )$ takes the minimal possible value, 
namely zero, at interior points (e.g., $F_{b^*,b}( -i \kappa e,  i \kappa e ) = 0$ for $\kappa >0$).
Consequently, $F_{b^*,b}( \bar  z_1 ,  z_1 )$ has to vanish identically. Inspecting the
r.h.s.\ of (42) we find that
\# { f_b^{(n)}   ( z_1) = 0 \qquad \forall  z_1 \in {\cal T} \quad \forall n \in \N.}
Consequently,
\# { {\cal F}_{b^*,b} (z_1, z_2) = 0 \qquad \forall (z_1,z_2) \in -{\cal T} \times {\cal T}.}
}

\Pr{(of Theorem 3.3).
Let $a, b \in \A$ with $ \| a \|=1$. According to Proposition 3.1 and
Theorem 3.2 there exists a sequence 
$\bigl\{ f_a^{(n) }\colon {\cal T} \to \C   \bigr\}_{n \in \n} $ of 
functions holomorphic in ${\cal T} $ which satisfies (35). Lemma 3.4 now allows us to
consider --- for $z \in {\cal T}$ and $a \in \A$ fixed --- the map
$\hat \phi_{a,z} \colon {\cal S} \to \C$ 
\# { \pi_\infty (b) \Omega_\infty  \mapsto 
\sum_{n = 1}^\infty \overline {  f_a^{(n)} ( z)} \, f_b^{(n)} (0) .}    
Here ${\cal S}$ denotes the set of vectors ${\cal S} := \{ \pi_\infty (b) \Omega_\beta : 
b \in \A \}$. Now $\overline{ \pi_\infty (\A) \Omega_\infty} = \H_\infty$ and
\# { \Bigl| \sum_{n \in \n} \overline {  f_a^{(n)} ( z)} \, f_b^{(n)} (0) \Bigr|^2 
\le F_{a^*,a} (\bar z, z) \cdot \| \pi_\infty (b) \Omega_\infty \|^2.} 
Thus the Hahn--Banach Theorem allows us to extend
the map $\hat \phi_{a,z} \colon {\cal S} \to \C$  to 
a (bounded) continuous linear functional
$\phi_{a,z}$ on $\H_\infty$. The Riesz Lemma ensures that
there exists a vector $\Phi_a (z) \in \H_\infty$ 
such that
\# { \phi_{a,z} (\Psi) = \bigl(\Phi_a (z)  \, , \, \Psi\bigr) \qquad \forall \Psi \in \H_\infty.}
As can easily be seen (by choosing once again an appropriate sequence $\{ a_n \}_{n \in \n}$
of analytic elements), the map
\# { z \mapsto \Phi_a (z) }
is analytic for $z \in {\cal T}$ and weakly continuous at the boundary set $\Im z = 0$,
where it satisfies
\# { \Phi_a (x) = \pi_\infty \bigl( \alpha_{x} (a) \bigr) \Omega_\infty  
\qquad \forall x \in \R^4.}  
}

Without proof we mention the following

\Cor{Let $a, b \in \A$ and let $\Phi_a $, $\Phi_b$ denote the associated vector valued functions
introduced in (48).
It follows that 
\# { {\cal F}_{a^*,b} (\bar z_1,z_2) = \bigl( \Phi_a (z_1)  \, , \, \Phi_b (z_2) \bigr) 
}
for all $z_1, z_2 \in {\cal T}$. Here ${\cal F}_{a^*,b}$ denotes the 
analytic function introduced in (10).}

What remains to be proven in order to establish the Reeh--Schlieder property
is fairly standard. Borchers and Buchholz [BoB] recently 
gave a nice and transparent formulation of this part of the argument and 
since it has already been reproduce in [J\"a], we feel free
to refer the reader to the literature cited. Thus we simply state our result. 

\Th{Consider a QFT as specified in Section 2 and let $\omega_\infty$ 
be a state, which satisfies the 
relativistic ground state condition. If the additivity assumption (22) holds, then
\# { \H_\infty = \overline { \pi_\infty \bigl( \A (\O) \bigr) \Omega_\infty},    }
for any open space--time region $ \O  \subset \R^4$. Moreover,
if the spacelike complement of $\O$ is not empty, then
$\Omega_\infty$ is separating for $ {\cal R}_\infty (\O )$.  } 

Similar to the situation in the vacuum and the KMS sector, 
$\Omega_\infty$  shares the Reeh--Schlieder property with 
a large class of vectors in $\H_\infty$. 

\Cor{Consider a QFT as specified in Section 2 and let $\omega_\infty$ 
be a state, which satisfies the 
relativistic ground state condition. Moreover, assume the additivity assumption (22) holds.
It follows that there exists a dense set ${\cal D}_\alpha \subset \H_\infty$,
namely
\# {  {\cal D}_\alpha = \Bigl\{ \Bigl( \1 - { \pi_\infty(a) \over 2 \, \| a \| } \Bigr) 
\Omega_\infty :  a \in \A_\alpha \Bigr\} ,}
such that for all $\Psi \in {\cal D}_\alpha$  
\# { \H_\infty = \overline { \pi_\infty \bigl( \A (\O) \bigr) \Psi},    }
where $ \O  \subset \R^4$ is again an arbitrary open space--time region.}

\Rem{The essential step is to show that for 
arbitrary $b \in \A$ the function
\# { \R^4 \ni x \mapsto \pi_\infty \bigl( \alpha_x (b) \bigr) \Psi }
extends to some analytic vector-valued function in the domain
${\cal T}$. The reader is invited to check that Proposition  3.1 and Theorem 
3.3 as well as Lemma 3.4 can easily be adapted and that the proofs given  
remain valid if we replace~$\Omega_\infty$ by some vector 
$\Psi \in {\cal D}_\alpha$.}

\vskip 1cm

\noindent
{\fourteenrm References}
\nobreak
\vskip .3cm
\nobreak
\halign{   &  \vtop { \parindent=0pt \hsize=33em
                            \strut  # \strut} \cr 
\REF
{BoB}
{Borchers, H.J.\ and Buchholz, D.}       {Global properties of vacuum states in de Sitter space}
                                  {Ann.\ l'Inst.\ H.\ Poincar\'e} 
                                                      {70} {23--40}
						                                                {1999}
\REF
{BB}
{Bros, J.\ and Buchholz, D.}      {Towards a relativistic KMS condition}
                                  {Nucl.\ Phys.\ B} 
                                  {429} {291--318}
                                  {1994}
% \REF
% {BB b}
% {Bros, J.\ and Buchholz, D.}      {Axiomatic analyticity properties and representations
%                                    of particles in thermal quantum field theory}
%                                   {Ann.\ Inst.\ H.\ Ponicar\'e} 
%                                   {64} {495--521}
%                                   {1996}
\REF
{BEM}
{Bros, J., Epstein, H.\ and Moschella, U.}      {Analyticity properties and thermal 
                                    effects for general quantum field theory on de Sitter
                                    space--time}
                                   {\CMP} 
                                   {196} {535--570}
                                   {1998}
\BOOK
{BR}  
{Bratteli, O.\ and Robinson, D.W.} {Operator Algebras and Quantum Statistical Mechanics~I, II} 
                                  {Sprin\-ger-Verlag, New York-Heidelberg-Berlin} 
                                  {1981}
\REF
{BJu}
{Buchholz, D.\ and Junglas, P.}   {On the existence of equilibrium states in local 
                                    quantum field theory} 
                                   {\CMP} 
                                   {121} {255--270}
                                   {1989}
% \REF
% {BJa}
% {Buchholz, D.\ and Jacobi, P.}     {On the nuclearity condition for massless fields} 
%                                    {\LMP} 
%                                    {13} {313--???}
%                                    {1987}
\HEP
{CH}
{Clifton, R.\ and Halvorson, H.}  {Generic Bell correlation between arbitrary local algebras
                                   in quantum field theory}
                                  {math-ph/9909013}
\BOOK
{G a}
{Glaser, V.}      {The positivity condition in momentum space}
                       {In {\sevensl Problems in Theoretical Physics.
                        Essays dedicated to N.N.\ Bogoliubov.} 
                        D.I. Bolkhintsev et al.\ eds.\ Moscow, Nauka}
                       {1969}
\REF
{G b}
{Glaser, V.}      {On the equivalence of the Euclidean and Wightman formulation of 
                   field theory}
                          {\CMP}	
                          {37}	{257--272}
                          {1974}
\BOOK
{H}
{Haag, R.}    {Local Quantum Physics: Fields, Particles, Algebras} 
              {Springer-Verlag, Berlin-Heidelberg-New York} 
              {1992}
% \REF
% {HHW}
% {Haag, R., Hugenholtz, N.M.\ and Winnink, M.}
%                           {On the equilibrium states in quantum statistical mechanics}  
%                           {\CMP}	
%                          {5}	{215--236}
%                          {1967}
% \REF
% {HKT-P}
% {Haag, R., Kastler, D.\ and Trych-Pohlmeyer, E.B.}    {Stability and equilibrium states}
%                                                      {\CMP} 
%                                                      {38} {173--193}
%						                                                {1974}
\REF
{J\"a}
{J\"akel, C.D.}                   {The Reeh--Schlieder property for thermal states}
                                  {\JMP}
				{41} {1--10} {2000}
% \BOOK
% {Ju}
% {Junglas, P.}   {Thermodynamisches Gleichgewicht und Energiespektrum in der 
%                 Quantenfeldtheorie} 
%                {Dissertation, Hamburg}
%                {1987}
% \REF
% {N}
% {Narnhofer, H.}   {Kommutative Automorphismen und Gleichgewichtszust\"ande}
%                  {Act.\ Phys.\ Austriaca}
%                  {47}   {1--29} 
%                  {1977}
%
% \REF
% {O}
% {Ojima, I.}   {Lorentz Invariance vs. Temperature in QFT}
%              {\LMP}
%              {11}    {73--80}
%              {1986}
% \REF
% {PW}
% {Pusz, W., and Woronowicz, S.L.}   {Passive states and KMS states for general 
%                quantum systems}
%               {\CMP}
%              {58}    {273--290}
%              {1978}
\BOOK
{R}
{Rudin, W.}     {Real and Complex Analysis} 
                {2nd ed.\ McGraw-Hill, New York}  
                {1974}
\REF
{RS}
{Reeh, H., and Schlieder, S.}     {Bemerkungen zur Unit\"ar\"aquivalenz von 
                                    Lorentzinvarianten Feldern}
                                   {Nuovo Cimemento} 
                                                      {22} {1051--}
						                                                {1961}
% \BOOK
% {S}
% {Sakai, S.}     {Operator Algebras in Dynamical Systems} 
%                 {Cambridge University Press, Cambridge-New York-Port Chester-Melbourne-Sydney}  
%                {1991}
\HEP
{St}
{Strohmaier, A.}                   {The Reeh--Schlieder property for the Dirac field 
                                    on static spacetimes}
                                   {math-ph/9911023} 
\REF
{SW}
{Summers, J.S., and Werner, R.F.} {On Bell's inequalities and quantum field theory. II.
                                   Bell's inequalities are maximally violated in the vacuum} 
                                  {\JMP} 
                                  {28}	{2440--2447}
                                  {1987}
\REF
{V}
{Verch, R.}                   {Antilocality and a Reeh--Schlieder theorem on manifolds}
                                   {\LMP} 
                                                      {28} {143--154}
						                                                {1993}
\cr}

\bye